# Localization and delocalization of charges injected in DNA

T. Heim[a], T. Mélin, D. Deresmes, and D. Vuillaume

Institut d'Electronique de Microélectronique et de Nanotechnologie,

CNRS - UMR 8520, BP 69 Avenue Poincaré, F-59652 Cedex Villeneuve d'Ascq, France

The electrical properties of DNA molecules are investigated by charge injection and electric force microscopy experiments. Prior to injection, DNA molecules exhibit a weak positively charged state. We probe the electrical behaviour of DNA by measuring the localized or delocalized character of the DNA charge states upon injection of excess charges. We show that injected charges do not delocalize for overstretched DNA prepared by a receding meniscus technique, while the adjunction of spermidine during the deposition leads to relaxed DNA molecules exhibiting a charge delocalization over microns. The interplay between charge localization/delocalization and deposition techniques may explain that transport behaviors ranging from insulating to conductive have been reported for DNA deposited on surfaces.

85.65.+h, 87.14.Gg, 87.64.Dz, 68.37.Ps

---

[a] Corresponding author : electronic mail : theim@isen.iemn.univ-lille1.fr

Charge transfer in DNA was first pointed out by Eley and Spevey in 1962 [1], after the discovery of the DNA double helix structure in 1953. The research on DNA conductivity is of interest as charge transfer seems to play a key-role in the protection of the genetic code [2]. Nevertheless, the experiments on the conductivity of DNA are far from having reached a consensus, with electrical behaviours ranging from insulator to metallic and even contact-induced superconductivity [2, 3]. This variety of experimental results may certainly be related to the variety of available DNA deposition methods. One of the most critical processes in DNA deposition is the drying step leading either to a condensed DNA phase, or to isolated molecules far from their natural environment. Recently, Kasumov and co-workers [4] have pointed out the interplay between the conformation of DNA molecules on surfaces and their electrical properties, based on height measurements by AFM and by using pentylamine surface treatments. The relationship between the DNA structure and its electrical properties was also established by Nakayama *et al.* [3], who studied the conductivity the DNA strands as a function of temperature, related to changes in the intramolecular stacking of base pairs close to the melting transition.

We investigate in the present Letter the charge localization and delocalization properties of DNA molecules as a function of their conformation on a substrate. The electrical behaviour of DNA is probed using local charge injection by the apex of an atomic force microscope tip, and by measuring the localized or delocalized character of charges injected in DNA by Electric Force Microscopy (EFM) experiments.

The samples are prepared as follows. Doped $n+$ silicon wafers with a ~2nm native oxide were cleaned in a piranha ($H_2SO_4$ : $H_2O_2$ 2:1) solution, leading to a highly hydrophilic OH – rich surfaces. To obtain hydrophobic surfaces for DNA deposition, wafers have been chemically treated with octadecyltrichlorosilane (OTS). The OTS molecules have been used as received from Petrarch, and dissolved ($10^{-3}$ – $10^{-1}$ M) in an organic solvent maintained at a constant temperature (20°C) in a dry nitrogen purged glove box. The freshly cleaned substrate was dipped into the solution for two hours. This silanization procedure is known to form a 2.5 nm thick self-assembled monolayer with good insulating properties [5]. The hydrophobic behaviour required to minimize grafting between DNA phosphates and the surface [4] was checked by contact angle measurements ($\theta_{water}$=108-110°, $\theta_{hexadecane}$=42°). In our experiments, indeed, DNA molecules exhibit at most a few grafting points with the surface (see Fig.1), and preferentially bind by the molecule ends [6].

DNA molecules of the λ phage (16 µm long – 48502 bp) were purchased from Roche – Biomedicals, dispersed in a TE buffer (Tris 10mM – EDTA 1mM) at pH = 6.5. The morphology of DNA deposition is changed using three methods. **In method (A)**, a drop of TE buffer containing DNA (at 250-500ng/ml, i.e. 10-20pM) was deposited on the OTS surface, left a few minutes for incubation under cover, and removed by tilting the sample. As seen in Fig. 1a, this method leads to overstretched ropes of a few (<5 typically) DNA molecules exhibiting an orientation perpendicular to the moving liquid – air meniscus similarly to Ref. [6]. The overstretching of DNA has been confirmed by fluorescence microscopy. **In method (B)**, the drop is left evaporating on the surface under cover. This leads to an increase of the incubation time (~5 hours), and of the TE and DNA concentrations within the drop during the evaporation. This technique results in ropes with bigger average size (ranging from a few to a few thousand molecules) and organized in a tree-shape, as seen from Fig. 1b. **In method (C)**, the ~ 40 µl drop of DNA in TE buffer is first deposited on the surface as in method (A). A ~1 µl droplet of spermidine (1M, pH~7), known as condensing agent[7], is then added to enhance the interaction between DNA molecules already grafted on the surface. After 30 minutes incubation under cover, the drop is removed by tilting the sample. Straight DNA ropes are not observed as in the overstretched case of method (A), but rather separated ropes of 5 to 20 nm in diameter showing no preferential orientation.

Charge injection and EFM experiments were performed using a Nanoscope III microscope (Digital Instruments) under dry nitrogen atmosphere. We used PtIr-coated cantilevers with frequency ~60 kHz, spring constant ~ 1-3 N/m. To inject charges into DNA, the EFM tip is biased at $V_{inj}$ with respect to the silicon wafer, and pressed with a typical 2nN contact force on a DNA rope for a few minutes. The resulting transfer of charges in DNA is then characterized by EFM, in which electric force gradients acting on the tip biased at $V_{EFM}$ shift the EFM cantilever phase [8]. EFM images reveal two distinct interactions. First, the capacitive interaction associated with the local increase of the tip-substrate capacitance when the EFM tip is moved over DNA molecules (see fig. 2, inset), leading to negative phase shifts varying as $-V_{EFM}^2$ (see dark features in Fig.2 a2-c2 associated with prominent topography features). Second, the interaction between the DNA charge $Q_{DNA}$ and capacitive charges at the tip apex. This additional phase shift is either positive or negative, and varies as $Q_{DNA}*V_{EFM}$. EFM images of charged DNA reported in Fig. 2 correspond to $Q_{DNA}*V_{EFM} > 0$, and thus to a positive phase shift, which can even overcome the capacitive DNA signals as seen from the bright features observed in Figs. 2a3-b3-c3-c4. In any case, capacitive and charge interactions can be distinguished in EFM measurements recorded as a function of $V_{EFM}$ [9].

Fig. 2a shows topographic AFM images of DNA ropes obtained using method (A) (Fig. 2a1, rope diameter ~4 nm), method (B) (Fig. 2b1, diameter ~10-20 nm) and method (C) (Fig. 2c1, diameter ~6-7nm). Corresponding EFM images prior to injection experiments are given in Fig. 2 a2, b2 and c2, respectively. The DNA ropes appear either as slightly dark (Fig.2a2) or slightly bright (Fig. 2b2) features in phase images. Bright features observed in the case of Fig. 2b2 at a positive tip bias ($V_{EFM}$) unambiguously indicate that DNA ropes are positively charged on the surface after drying. For the three methods, charge and capacitive effects were carefully separated, revealing that DNA ropes are positively charged on the surface *in the three cases*. From the ratio between capacitive and charge EFM signals, we estimate the effective volume charge of DNA ropes using a model derived from Ref. [9]: in case of method (A), an effective charge density of 1.1 (± 0.2) $10^{18}$ cm$^{-3}$ (*i.e.* one charge every 1000 base pairs) is found. For method (B) we found 3.0 (± 0.5) $10^{17}$ cm$^{-3}$ (*i.e.* one charge every 3000bp), and for method (C) we found 7.2 (± 5) $10^{17}$ cm$^{-3}$ (i.e. one charge every 1500bp). This result is striking since DNA is expected to be negatively charged in solution even in presence of counter-ions screening its intrinsic charge (2 electrons per bp). The explanation of the observed positive charge of DNA falls beyond the scope of this paper, but may originate in tunneling of DNA electrons through the ultrathin OTS and SiO$_2$ layers to the underlying conductive substrate, as observed *e.g.* in the case of CdSe nanocrystals on conductors [10].

In Fig. 2, we showed the EFM data of DNA ropes after injection of excess charges. For deposition methods A and B, injected charges spread at most over a few hundreds of nm, as seen from the bright EFM features. This order of magnitude of charge spreading (see Fig. 2-a3) is typical of charge injection in a dielectric insulating layer (here the OTS and SiO$_2$ layers) and is due to a non-zero in-plane electric field induced by the tip during the injection [11]. In the case method B (Fig. 2-b3), the elongated EFM signal also reveals a faint charge transfer over ~500 nm along the DNA rope . The case of deposition method (C) is striking, since excess charges are seen to delocalize *over the whole length of the DNA ropes*, *i.e.* over several micrometers. As measured from Fig. 2c3, the amount of stored charges is ~3 times greater than before injection, and shows a typical 1 hour charge decay time scale.

We moreover proved able to modify the charge delocalization properties of a given rope. This is illustrated in Fig. 2-c4 showing an electrical breakdown at the point denoted by a circle. The electrical breakdown was created by an injection experiment at fairly high voltage (here using $V_{inj}$=-6V for 4 minutes, after experiment of fig. 2-c3 and before experiment of fig. 2-c4). The charge injection and EFM experiments of fig.2-c4 were performed in similar conditions as in Fig.2-c3.

In the following we interpret our experimental results using the framework of large-distance conduction mediated by electronic coupling between the localized molecular orbitals of the DNA base pairs[2]. Ionic conduction is unlikely, as the hydration shell surrounding DNA is reduced under dry nitrogen conditions [12].

In the case of method (A), the overstretching of DNA [6] results in a loss of stacking and coupling between base pairs resulting in insulating properties[2]. Though DNA is not likely to be overstretched in the case of method (B), its insulating behaviour may arise from a structural transition from the B to A form [13], which has been shown to occur for DNA fibers under dry nitrogen, and corresponds to vanishing electronic couplings due to the twist angle variation from the B to the A-form as predicted by DFT calculation [2].

The long distance charge spreading observed in the case of method (C) shows that the adjunction of spermidine during DNA deposition strongly affects the properties of charge transfer. Spermidine is known as a condensing agent which stabilize the B-DNA structure within a rope: its positively charged amine groups make contact with the negatively charged phosphates of adjacent DNA helices, thus constraining DNA in its B–form even under dry nitrogen atmosphere [14], resulting in a possible charge transfer along the DNA rope. The electrical breakdown observed in Fig. 2c4 with no detectable alteration in topography shows that the electrical pathway in DNA can be also stopped without mechanical cutting of the rope, suggesting that the charge delocalization is indeed related to the internal structure of the DNA rope.

We showed that injection and charge delocalization measured by EFM is a powerful tool to study the electronic behaviour of DNA molecules. Upon condensation using spermidine, DNA molecules display a drastic enhancement of the spreading of injected excess charges, suggesting that the localization/delocalization of charges is related to the conformation of DNA molecules on the surfaces and the coupling between the base pair molecular orbitals. This charge injection scheme may be generalized to study the local action of proteins, enzymes, or local denaturation on the conduction pathways in DNA.

The authors thank H.Bouchiat, E. Le Cam, O.Pietrement, E.Delain and R. Blossey for fruitful discussions, and acknowledge a partial financial support by IRCICA.

FIGURE CAPTIONS

FIG. 1. AFM scans of λ-phage DNA prepared in a TE buffer (Tris 10mM – EDTA 1mM) at pH = 6.5, and deposited on a self-assembled monolayer of octadecyltrichlorosilane (OTS) on a silicon wafer. a) 5x5 µm² AFM scan (10 nm z-scale) of DNA following the deposition method (A) (see inset), in which the 40 µl drop of DNA in TE is left incubating under cover (*i*) for a few minutes, and removed by tilting the sample (*ii*). The arrow in the AFM image indicates the receding meniscus direction. b) 10x10 µm² scan (40 nm z-scale) following method (B): the 40 µl drop of DNA in TE is left drying during 5 hours under cover (*i* and *ii*), leading to an organization of DNA ropes in a tree shape. The arrows indicate the direction of receding meniscus during drying. c) 7x7 µm² AFM scan (20 nm z-scale) following method (C), in which the 40 µl drop of DNA is first deposited on the OTS surface, followed by adding a 1 µl droplet of spermidine (*i*). After 30 minutes incubation (*ii*), the solution is removed by tilting the sample (*iii*). The arrow in the AFM image indicates the receding meniscus direction

FIG.2. Inset: schematics of the charge injection (top), in which the EFM tip is brought in contact (applied force ~2nN) on a DNA rope, and polarized at $V_{inj}$ with respect to the substrate. The charge detection (bottom) is obtained in the electric force microscopy mode, in which one moves the tip in a linear pass at a distance z~60-90 nm from the sample plane, and one records the oscillating cantilever phase shift as a probe of electrical force gradients (see text). a1), b1) and c1): topography images of DNA

deposited using methods (A), (B) and (C). The respective scan sizes and z-scales are 2.5x1.25 µm² and 10 nm, 3.6x3.15 µm² and 50 nm, and 3x3µm² and 30 nm. a2), b2), c2): corresponding EFM phase images prior to injection experiments (4°, 2° and 2° phase colour-scales, respectively). EFM tip voltages are: $V_{EFM}$=+5V, $V_{EFM}$=+4V, and $V_{EFM}$=+6V, respectively. a3), b3), c3): corresponding EFM images after charge injection. Injection points are denoted by arrows. The injections have been carried out using a $V_{inj}$=+4V tip bias and injection times: 2 min, 4 min, and 4 min, respectively. EFM phase images have been acquired in the same experimental conditions as in a2), b2) and c2). The phase colour-scale is 2° for all. c4) EFM image corresponding to c3) after injection at $V_{inj}$=+4V during 4 minutes. The injection point is denoted by an arrow. The electrical breakdown observed at the point denoted by a circle results from a previous injection experiment.

**Figure 1**  (T. Heim, T. Mélin, D. Deresmes and D. Vuillaume)

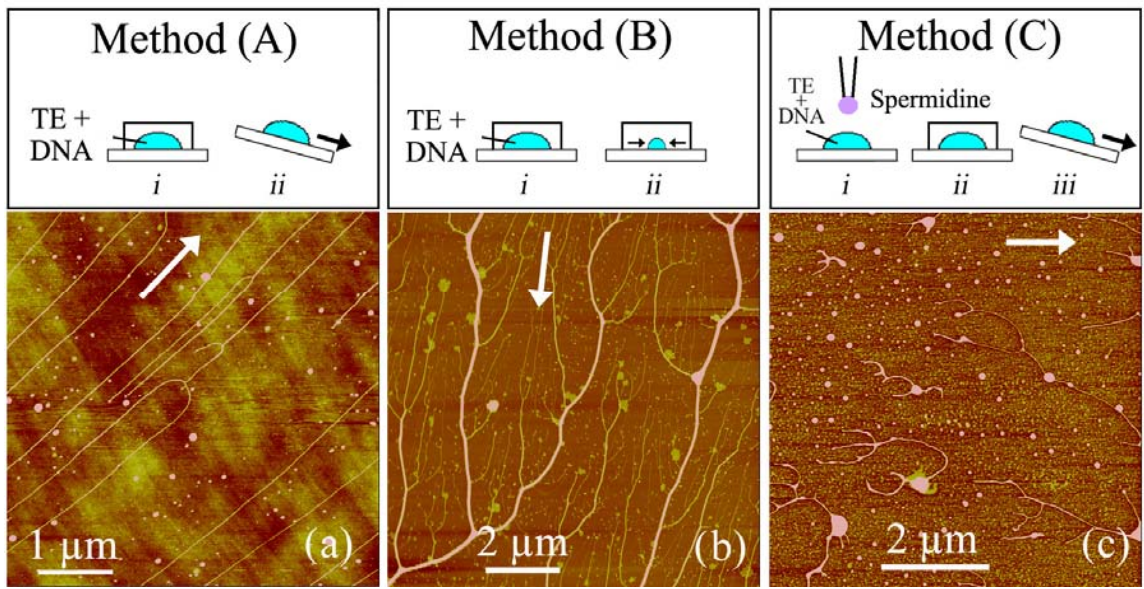

**Figure 2**  (T. Heim, T. Mélin, D. Deresmes and D. Vuillaume)

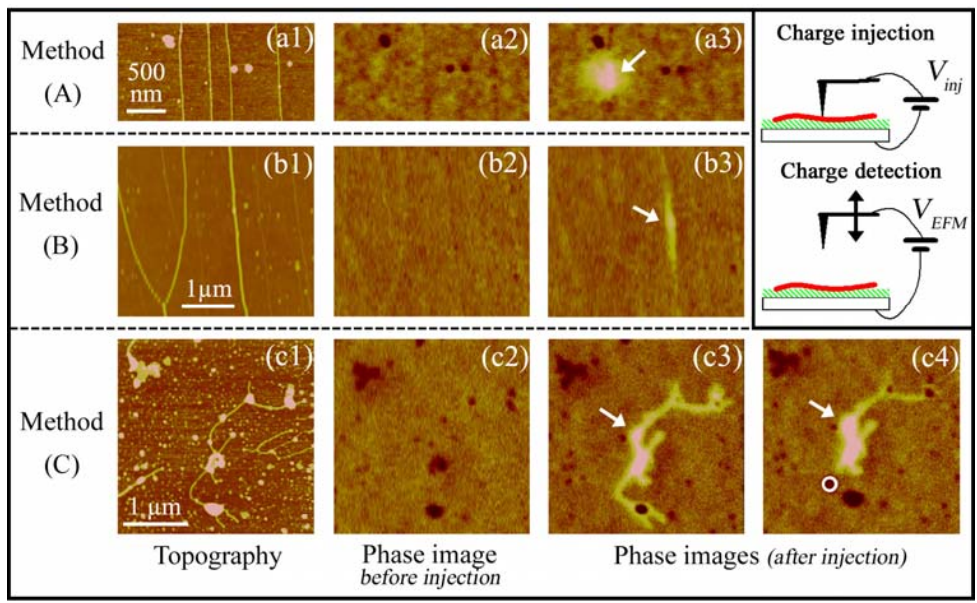